\documentclass[useAMS,usenatbib]{mn2e}
\usepackage{graphicx}
\usepackage{txfonts}

\title[SDSS galaxies with double-peaked emission lines]
      {SDSS galaxies with double-peaked emission lines: double starbursts or AGNs?}

\author[L.S.Pilyugin et al.]
       {L.S.~Pilyugin$^{1}$,
        I.A.~Zinchenko$^{1}$,
        B.~Cedr\'{e}s$^{2,3}$,  
        J.~Cepa$^{2,3}$,  
        A.~Bongiovanni$^{2,3}$,
        L.~Mattsson$^{4}$,  
    \newauthor  J.M.~V\'{\i}lchez$^{5}$  \\
     $^{1}$ Main Astronomical Observatory
            of National Academy of Sciences of Ukraine,
            27 Zabolotnogo str., 03680 Kiev, Ukraine \\
     $^{2}$ Instituto de Astrof\'{\i}sica de Canarias, C/v\'{i}a L\'{a}ctea S/N,
            38200 La Laguna, Spain \\
     $^{3}$ Departamento de Astrof\'{\i}sica, Universidad de La Laguna,E-38071 La Laguna, Tenerife, Spain \\ 
     $^{4}$ Dark Cosmology Centre, Niels Bohr Institute,
            University of Copenhagen, Juliane Maries Vej 30,
            DK-2100, Copenhagen \O, Denmark\\
     $^{5}$ Instituto de Astrof\'{\i}sica de Andaluc\'{\i}a,
            CSIC, Apdo, 3004, 18080 Granada, Spain \\
             }

\date{Accepted 2011 August 26. Received 2011 August 03; in original form 2011 March 23}

\begin{document}

\maketitle

\begin{abstract}
With the aim of investigating galaxies with two strong simultaneous starbursts, 
we have extracted a sample of galaxies  with double-peaked emission lines in their global 
spectra from the SDSS spectral database. 
 We then fitted the emission lines H${\alpha}$, H${\beta}$, [O\,{\sc iii}]$\lambda$5007, 
[N\,{\sc ii}]$\lambda$6584, [S\,{\sc ii}]$\lambda$6717 and [S\,{\sc ii}]$\lambda$6731 
of 129 spectra by two Gaussians 
to separate the radiation of the two (blue and red) components. A more or less reliable 
decomposition of the all those emission lines have been found for 55 spectra.
Using a standard BPT classification diagram, we have been able to divide the galaxies from our sample 
into two subsamples: Sample A consisting  of 18 galaxies where both components 
belong to the photoionised class of objects, and Sample B containing 37 galaxies 
which show non-thermal ionisation (AGNs).
We have examined the properties of the blue and red components, and found that 
the differences between radial velocities of components lie within 200 -- 400 
km s$^{-1}$ for galaxies of both subsamples. 
The equivalent number of ionising stars is in the range 10$^4$ -- 10$^5$ O7V stars for 
each component in the galaxies of Sample A. 
We have estimated the oxygen and nitrogen abundances as well as the electron 
temperatures for each component using the recent NS-calibration and from 
global spectra for galaxies from Sample A using both the NS and ON-calibration.  
We have found that the global oxygen abundance is typically in between the measured abundances 
of individual components for our sample of galaxies, and that both calibrations provide consistent 
global abundances. 
Finally, we suggest the classical O/H -- N/O diagram is used to test the reliability of 
the dividing lines between starburst-like objects and AGNs in the so-called BPT diagram.  
\end{abstract}


\section{Introduction}

The study of starburst galaxies is very important for understanding 
both star formation and (chemical) evolution of galaxies. 
The Sloan Digital Sky Survey \citep[SDSS,][]{yorketal2000}
provides a very large database of spectra of galaxies,
which has been used in many studies of the chemical evolution of galaxies 
\citep [see, e.g.,][]{kniazevetal2004,izotovetal2004,tremontietal2004,thuanetal2010,pilyuginthuan2007,pilyuginthuan2011}.

The SDSS spectra are obtained through 3-arcsec diameter fibers. 
At a redshift of $z=0.12$ (which is a mean value of the redshifts of galaxies 
considered at the present study), the projected aperture diameter is $\sim$ 7 kpc. 
This suggests that the SDSS spectra of the considered galaxies are closer to global spectra 
(composite nebulae that include multiple star clusters), rather than to 
spectra of individual H\,{\sc ii} regions. 
In a typical case,  many individual H\,{\sc ii} regions are distributed over 
the disc of a galaxy. Due to the rotation of galaxies, H\,{\sc ii} regions 
will have different radial velocities depending on the inclination of their host galaxy.
One may expect that the emission line profile in the global SDSS spectra 
of such composite nebulae in distant galaxies can be described by a Gaussian which 
is wider than that for an individual H\,{\sc ii} region. 
If two strong starbursts take place in a galaxy and the giant H\,{\sc ii} regions 
associated with those starbursts make a dominant contribution to the radiation 
in the emission lines  then one could expect 
double--peaked emission line profile in the global spectrum of a galaxy. 
While double-peaked emission lines have been extensively studied in AGN and radio galaxies,
double starburst galaxies are yet unexplored. Those systems may induce 
errors in abundance determinations due to the mixing of two starburst events. Then, establishing 
the abundance of those double-peaked starburst abundances as global systems, 
with respect to that defined by their components, will shed light on the role and possible impact 
of those double-peaked emission line galaxies on abundance determinations.
To this aim, we have carried out a search for SDSS spectra of galaxies with double-peaked  emission 
line profiles, revealing several hundreds of candidates where two strong 
starbursts can be observed simultaneously. 

\begin{figure*}
\resizebox{1.00\hsize}{!}{\includegraphics[angle=000]{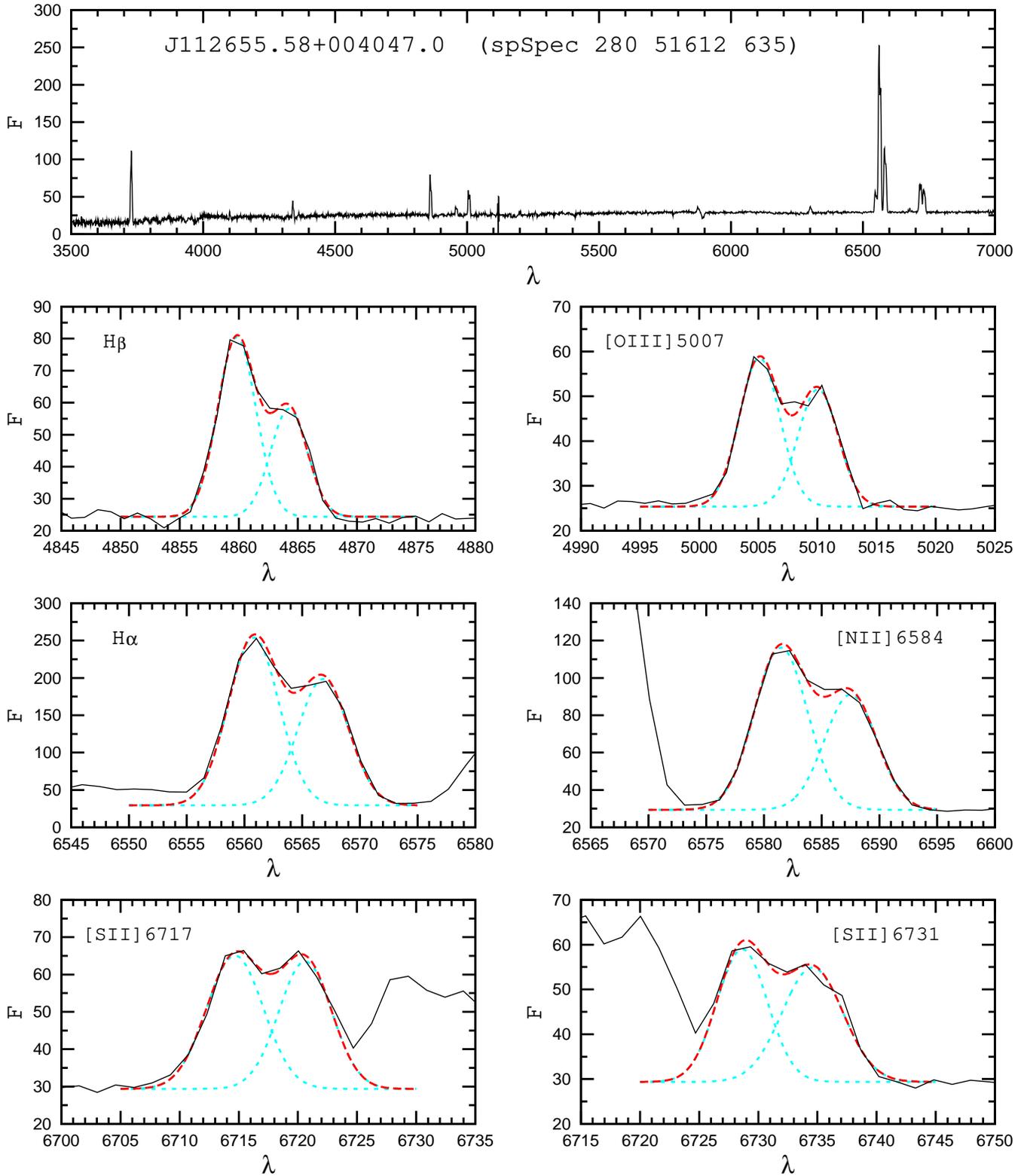}}
\caption{
The observed spectrum of the SDSS object J112655.58+004047.0 (spSpec 280 51612 635) 
and double-Gaussian fits to the emission lines  
H${\beta}$, [O\,{\sc iii}]$\lambda$5007, H${\alpha}$, 
[N\,{\sc ii}]$\lambda$6584, [S\,{\sc ii}]$\lambda$6717 and [S\,{\sc ii}]$\lambda$6731. 
The solid line is the observed line profile. The long dashed (blue) lines are 
the profiles of the blue and red components, respectively, and the short dashed 
(red) line is the sum of the blue and red components.
(A color version of this figure is available in the online version.)
}
\label{figure:fit}
\end{figure*}

Here, we separate the radiation of two giant H\,{\sc ii} regions 
associated with two different starbursts by the fitting two Gaussians to the emission lines 
in the global spectra. We examine the properties of each component of these "binary starbursts": 
the electron density, the oxygen and nitrogen abundances, and the number of the ionising 
stars they contain. These objects provide also an additional possibility to test how
representative abundances derived from global spectra are for the whole galaxies. 
Comparison between oxygen abundances in the blue and red components and 
global oxygen abundances will tell us something about the reliability
of the abundances derived from global spectra.  We test the reliability of global abundances 
by comparison of global abundances obtained using two recent strong-line calibrations as well.

Finally, we suggest to use the classical O/H -- N/O diagram to test of the credibility of 
the dividing lines between the starburst-like objects and the AGNs in the \citet{baldwinetal1981} 
(BPT) classification diagram.  

The paper is organized as follows.
Decomposition of the double-peaked emission line profiles in the global SDSS spectra 
and the selection of galaxies are described in Section 2.
The general properties of the selected sample of galaxies are given in Section 3. 
The oxygen and nitrogen abundances and the electron temperatures derived for 
individual (blue or red) components and from global spectra are discussed in Section 4 and
Section 5 gives a brief summary of the main results.  

Throughout the paper, we will be using the following notations for the line fluxes,
\begin{equation}
R_2 =  I_{\rm \rm [OII] \lambda 3727+ \lambda 3729} /I_{\rm {\rm H}\beta },
\end{equation}
\begin{equation}
N_2  = I_{\rm \rm [NII] \lambda 6548+ \lambda 6584} /I_{\rm {\rm H}\beta },
\label{equation:n2standard}
\end{equation}
\begin{equation}
S_2  = I_{\rm \rm [SII] \lambda 6717 + \lambda 6731} /I_{\rm {\rm H}\beta },
\end{equation}
\begin{equation}
R_3  = I_{\rm {\rm [OIII]} \lambda 4959 + \lambda 5007} /I_{\rm {\rm H}\beta }.
\label{equation:r3standard}
\end{equation}
 The [O\,{\sc iii}]$\lambda$5007 and $\lambda$4959 lines originate from transitions from the 
same energy level, so their fluxes ratio is due only to the transition probability ratio 
which is close to 3 \citep{storey2000}. Hence, the value of $R_3$ can be well approximated by    
\begin{equation}
R_3     = 1.33 \; I_{\rm {\rm [OIII]} \lambda 5007} /I_{\rm {\rm H}\beta }.
\label{equation:r3alternat}
\end{equation}
Similarly, the [N\,{\sc ii}]$\lambda$6584 and $\lambda$6548 lines also originate from transitions from the 
same energy level and the transition probability ratio for those lines is again
close to 3 \citep{storey2000}. The value of $N_2$ is therefore well approximated by
\begin{equation}
N_2  = 1.33 \; I_{\rm \rm [NII] \lambda 6584} /I_{\rm {\rm H}\beta },
\label{equation:n2alternat}
\end{equation}
The electron temperatures $t$ are given in units of 10$^4$K.

\section{THE DATA}

As the first step, we have extracted several hundred spectra of galaxies 
with double-peaked emission lines from the data release 7  \citep{abazajianetal2009}  
of the Sloan Digital Sky Survey (SDSS). 
In our previous work we have visually inspected 
SDSS spectra aiming to search for spectra where the two auroral lines [O\,{\sc iii}]$\lambda$4363 and 
[N\,{\sc ii}]$\lambda$5755 can be detected simultaneously \citep{pilyuginmnras2010}. 
We have also collected spectra  with double-peaked emission lines. Although we have considered all 
the SDSS DR7 spectra, we do not suggest our sample of SDSS spectra with double-peaked emission 
lines is exhaustive. 

In order to separate the radiation of the blue and red components, we have fitted two Gaussians to 
the emission lines (H${\alpha}$, H${\beta}$, [O\,{\sc iii}]$\lambda$5007, 
[N\,{\sc ii}]$\lambda$6584, [S\,{\sc ii}]$\lambda$6717 and [S\,{\sc ii}]$\lambda$6731) 
in the extracted global spectra. 
The contribution of each individual component to the global flux have been derived
as illustrated by the following example, concerning the H$\beta$ line.
The continuum flux f$_c$ is assumed to be constant between $\lambda_a = 4836$\AA $ $ and  
$\lambda_b = 4886$\AA $ $ (all spectra have been shifted to zero redshift).
The line profiles of the blue and red components are approximated by Gaussians, i.e.,
\begin{equation}
f(\lambda)= F\, \frac{1}{\sqrt{2\pi}\,\sigma} 
e^{-(\lambda - \lambda_0)^2/2\sigma^2}
\label{equation:gauss}
\end{equation}
where $\lambda_0$ is the central line wavelength, $\sigma$ is the width of the line, and 
$F$ is the flux in the emission line.
Thus,  the total flux at a fixed value of $\lambda$ is given by the expression
\begin{equation}
f(\lambda)= f_{\rm H\beta, \,{\rm blue}}(\lambda) + f_{\rm H\beta, \,{\rm red}}(\lambda) + f_{\rm \rm c}(\lambda).
\label{equation:ftot}
\end{equation}

For the Gaussian fitting, we have used the robust non-linear least squares curve 
fitting package MPFIT\footnote{http://purl.com/net/mpfit} described in \citet{markwardt2009}.
MPFIT works much faster then a "brute-force" approach, but in some cases it may 
only find a local minimum. 
To avoid this, we calculated a set of best-fit parameters using MPFIT. 
We have considered $\Delta \lambda = \lambda_0^{red}$ - $\lambda_0^{blue}$ on the interval corresponding 
to differences between radial velocities of the blue and red components 
from 120 to 600 km s$^{-1}$. This range was divided into 160 intervals $\Delta \lambda_i$ 
with step 3  km s$^{-1}$. 
The best-fit parameters for every interval of  $\Delta \lambda_i$   
are derived by finding the minimum of the mean difference, 
\begin{equation}
\epsilon = \sqrt{ \frac{1}{n} \sum_{\rm k=1}^{k=n} (f(\lambda_k)-f^{obs}(\lambda_k))^2},
\label{equation:df}
\end{equation}
between the measured flux f$^{\rm obs}(\lambda_k)$ and 
the flux f($\lambda_k$) given by Eq.(\ref{equation:ftot}) 
on the wavelength interval between $\lambda_a$ and $\lambda_b$.
The "true" values of 
F$(H{\beta}, \,{\rm blue})$, $\lambda_0(H{\beta}, \,{\rm blue})$, $\sigma(H{\beta}, \,{\rm blue})$, 
F$(H{\beta},\,{\rm red})$, $\lambda_0(H{\beta},\,{\rm red})$, $\sigma(H{\beta}, \,{\rm red})$
 are then selected by comparing the minimum values of $\epsilon_i$ for different intervals of 
 $\Delta$$\lambda$$_i$. 
In other words, we have used a two-step procedure to obtain the best-fit parameters.

 The fit to the  [O\,{\sc iii}]$\lambda$5007 line is obtained using the wavelength interval 
from $\lambda_a$ = 4982 \AA $ $  to $\lambda_b$ = 5032 \AA. 
The continuum flux f$_c$ is assumed to be constant between
$\lambda_a$ = 6523 \AA  $ $  and  $\lambda_b$ = 6608 \AA $ $  in fitting of the 
 H$_{\rm \alpha}$ and [N\,{\sc ii}]$\lambda$6584 lines, and  between
$\lambda_a$ = 6690 \AA  $ $  and  $\lambda_b$ = 6760 \AA $ $  in fitting of the 
 [S\,{\sc ii}]$\lambda$6717 and [S\,{\sc ii}]$\lambda$6731 lines. 
The fits to the H${\alpha}$, H${\beta}$, [O\,{\sc iii}]$\lambda$5007, 
[N\,{\sc ii}]$\lambda$6584, [S\,{\sc ii}]$\lambda$6717 and [S\,{\sc ii}]$\lambda$6731 lines 
in the spectra (spSpec 280 51612 635) of the SDSS object J112655.58+004047.0 
are shown in Fig.~\ref{figure:fit}.

We have found 129 spectra where a double-Gaussian fit can be obtained for each emission line.
Unfortunately the value of the parameter  $\epsilon$  given by Eq.(\ref{equation:df}) 
cannot serve as an effective criterion for selecting spectra with reliable decompositions.  
First of all, the value of $\epsilon$ is a measure of the quality of the 
double-Gaussian fit to the global line profiles rather than a measure of 
the quality of the decomposition. In addition, the observed double-peaked line profiles  
are covered by about ten wavelength points only at the SDSS spectroscopic resolution. 
Hence, the value of $\epsilon$ may be affected significantly by the noise from just one data point. 
Therefore we have selected spectra where the 
decomposition of the blue and red components for each emission line
seem to be reliable, based upon visual inspection of the double-Gaussian fits (how well a 
double-peak is seen in each emission line), taking into account the limited quality of the SDSS spectra.
Our final list contains 55 spectra. Certainly, our selection is somewhat subjective, but
as for the spectra excluded from further consideration here,
the only lines that cannot be reliable 
decomposed into two components are [S\,{\sc ii}]$\lambda$6717 or/and [S\,{\sc ii}]$\lambda$6731.
The other lines can be fitted rather well by two Gaussians. 
Even if the [S\,{\sc ii}]$\lambda$6717 and [S\,{\sc ii}]$\lambda$6731 lines will not 
be used as separate lines in the most part of this study,
we have used the fits to these lines as a test of the robustness of the sample.

Since measurements of the [N\,{\sc ii}]$\lambda$6584 and [O\,{\sc iii}]$\lambda$5007 lines are 
more reliable than those of  the [N\,{\sc ii}]$\lambda$6548 and [O\,{\sc ii}]$\lambda$4959 lines, 
we have used  Eq.(\ref{equation:n2alternat}) to obtain the value of  
$N_2$ and  Eq.(\ref{equation:r3alternat}) to determine  the value of $R_3$.

The measured emission fluxes $F$ have been corrected for interstellar reddening.  
We have obtained the extinction coefficient C(H$\beta$)  
using the theoretical H$\alpha$ to H$\beta$ ratio ($= 2.86$) and the analytical 
approximation to the Whitford interstellar reddening law by \citet{izotovetal1994}. 

The redshift $z$ and stellar mass $M_{\rm \rm S}$ of each galaxy were taken from 
the MPA/JHU catalogs \footnote{The catalogs are available at 
http://www.mpa-garching.mpg.de/SDSS/}.
The techniques used to construct the catalogues are described 
in \citet{brinchmannetal2004,tremontietal2004} and other publications of those authors.

\section{General properties of the selected starbursts}

\subsection{BPT classification diagram}

\begin{table*}
\caption[]{\label{table:list}
Oxygen and nitrogen abundances and electron temperatures of individual components 
as well global values of Sample A. }
\begin{center}
\begin{tabular}{cccccccc} \hline \hline
SDSS number$^a$                           & 
Spectrum number$^b$                       & 
12+log(O/H)$^{blue}_{\rm NS}$                   & 
12+log(O/H)$^{red}_{\rm NS}$                    & 
12+log(N/H)$^{blue}_{\rm NS}$                   & 
12+log(N/H)$^{red}_{\rm NS}$                    & 
t$^{blue}_{\rm NS}$                             & 
t$^{red}_{\rm NS}$                              \\[3mm]
                                          & 
                                          & 
12+log(O/H)$^{glob}_{\rm NS}$                   & 
12+log(O/H)$^{glob}_{\rm ON}$                   & 
12+log(N/H)$^{glob}_{\rm NS}$                   & 
12+log(N/H)$^{glob}_{\rm ON}$                   & 
t$^{glob}_{\rm NS}$                             & 
t$^{glob}_{\rm ON}$                             \\  \hline
J015847.30-101802.7 &  665  52168 018 &  8.41 &  8.45 &  7.44 &  7.37 &  0.89 &  0.87  \\
                    &                 &  8.43 &  8.45 &  7.41 &  7.45 &  0.88 &  0.86  \\  
J084845.58+542329.1 &  446  51899 595 &  8.47 &  8.24 &  7.69 &  6.85 &  0.80 &  1.06  \\
                    &                 &  8.41 &  8.44 &  7.47 &  7.55 &  0.90 &  0.87  \\
J091911.35+531424.0 &  553  51999 639 &  8.48 &  8.50 &  7.67 &  7.57 &  0.80 &  0.79  \\
                    &                 &  8.49 &  8.46 &  7.63 &  7.46 &  0.80 &  0.83  \\
J103007.07+412353.5 & 1360  53033 186 &  8.30 &  8.41 &  6.95 &  7.15 &  1.00 &  0.96  \\  
                    &                 &  8.37 &  8.42 &  7.07 &  7.25 &  0.97 &  0.95  \\   
J103404.17+061210.2 &  999  52636 150 &  8.22 &  8.16 &  6.85 &  6.71 &  1.09 &  1.12  \\
                    &                 &  8.20 &  8.24 &  6.77 &  6.89 &  1.11 &  1.09  \\   
J103822.61+505814.7 & 1009  52644 042 &  8.48 &  8.42 &  7.70 &  7.47 &  0.80 &  0.87  \\  
                    &                 &  8.45 &  8.45 &  7.59 &  7.55 &  0.83 &  0.84  \\   
J104137.27+103252.0 & 1600  53090 020 &  8.52 &  8.61 &  7.98 &  7.97 &  0.74 &  0.69  \\  
                    &                 &  8.57 &  8.57 &  7.99 &  7.97 &  0.71 &  0.72  \\  
J112655.58+004047.0 &  280  51612 635 &  8.49 &  8.42 &  7.73 &  7.53 &  0.78 &  0.86  \\
                    &                 &  8.46 &  8.44 &  7.64 &  7.55 &  0.82 &  0.83  \\  
J113122.19-005606.4 &  281  51614 007 &  8.41 &  8.37 &  7.47 &  7.20 &  0.89 &  0.93  \\
                    &                 &  8.39 &  8.41 &  7.33 &  7.36 &  0.91 &  0.89  \\ 
J113413.84+533601.2 & 1014  52707 096 &  8.41 &  8.15 &  7.31 &  6.70 &  0.92 &  1.12  \\
                    &                 &  8.41 &  8.45 &  7.15 &  7.23 &  0.94 &  0.95  \\
J113515.65+234606.0 & 2501  54084 010 &  8.40 &  8.32 &  7.12 &  6.93 &  0.87 &  0.99  \\  
                    &                 &  8.39 &  8.45 &  7.09 &  7.18 &  0.91 &  0.90  \\  
J113835.63+281801.1 & 2220  53795 089 &  8.41 &  8.33 &  7.42 &  6.95 &  0.89 &  0.90  \\
                    &                 &  8.41 &  8.41 &  7.32 &  7.26 &  0.90 &  0.90  \\  
J114234.66-030527.9 &  329  52056 320 &  8.42 &  8.33 &  7.38 &  6.97 &  0.87 &  0.97  \\  
                    &                 &  8.37 &  8.43 &  7.06 &  7.13 &  0.93 &  0.93  \\  
J124844.42+250522.6 & 2661  54505 412 &  8.43 &  8.55 &  7.59 &  8.01 &  0.83 &  0.72  \\  
                    &                 &  8.47 &  8.44 &  7.73 &  7.58 &  0.80 &  0.82  \\  
J132253.24+585942.2 &  959  52411 397 &  8.46 &  8.54 &  7.64 &  7.86 &  0.82 &  0.73  \\  
                    &                 &  8.50 &  8.47 &  7.76 &  7.59 &  0.77 &  0.80  \\  
J141439.58+033601.8 &  583  52055 137 &  8.38 &  8.47 &  7.44 &  7.66 &  0.89 &  0.80  \\  
                    &                 &  8.42 &  8.44 &  7.54 &  7.56 &  0.84 &  0.83  \\  
J142504.46+050556.7 &  584  52049 571 &  8.45 &  8.43 &  7.63 &  7.44 &  0.82 &  0.88  \\
                    &                 &  8.44 &  8.42 &  7.54 &  7.45 &  0.85 &  0.87  \\
J161555.12+420624.5 & 1171  52753 376 &  8.51 &  8.38 &  7.40 &  7.08 &  0.89 &  0.96  \\  
                    &                 &  8.48 &  8.50 &  7.33 &  7.45 &  0.91 &  0.90  \\  
\hline 
\end{tabular}\\
\end{center}
\begin{flushleft}
$^a$ The objects are listed in order of right ascension. \\
$^b$ The spectrum number is composed of the SDSS plate number, 
the modified Julian date of the observation, 
and the number of the fiber on the plate.
\end{flushleft}
\end{table*}

\begin{figure}
\resizebox{1.00\hsize}{!}{\includegraphics[angle=000]{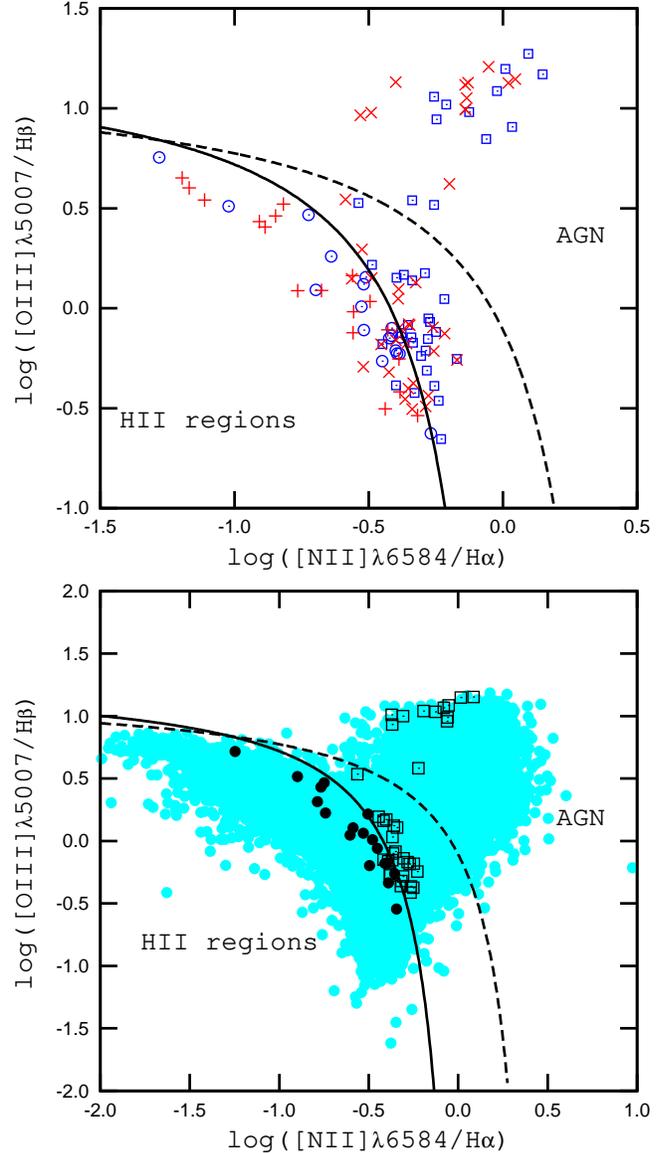}}
\caption{
The [N\,{\sc ii}]$\lambda$6584/H$\alpha$ versus [O\,{\sc iii}]$\lambda$5007/H$\beta$ diagram. 
{\it Upper  panel.} The open (blue) circles are the blue components and the (red) plus signs are 
the red components of the photoionised objects.
The open (blue) squares are the blue components and the (red) crosses are the
red components of the AGNs.
The solid line separates objects with H\,{\sc ii} spectra from those 
containing an AGN according to \citet{kauffmannetal2003}, while the dashed line is the same
according to \citet{kewleyetal2001}. 
{\it Lower  panel.} 
The gray (light-blue) filled circles show a large sample of emission-line SDSS 
galaxies from \citet{thuanetal2010}. 
The dark (black) filled circles show our sample of SDSS galaxies with 
double-peaked emission lines (global spectra, sum of blue and red components).
(A color version of this figure is available in the online version.)
}
\label{figure:seagull}
\end{figure}

The intensities of strong, easily measured lines can be used to separate  
different types of emission-line objects according to their main  
excitation mechanism (i.e. starburst or AGN). \citet{baldwinetal1981}  proposed a diagram 
(BPT classification diagram) where 
the excitation properties of H\,{\sc ii} regions are studied by 
plotting the low-excitation [N\,{\sc ii}]$\lambda$6584/H$\alpha$ 
line ratio against the high-excitation [O\,{\sc iii}]$\lambda$5007/H$\beta$ line ratio. 
Using the BPT classification  diagram, we have divided all galaxies from our sample 
into two subsamples. 
The first subsample (A) consists of 18 galaxies where both components (blue and red) 
are starburst like objects. 
The galaxies in Sample A are listed in Table~\ref{table:list}. 
The second subsample (B)  contains 37 galaxies where one or both components  show non-thermal ionisation
(here referred to as AGNs).  It should be noted that all components, in all galaxies, in our sample have narrow 
lines regardless of their positions in the 
 [N\,{\sc ii}]$\lambda$6584/H$\alpha$ versus [O\,{\sc iii}]$\lambda$5007/H$\beta$ diagram. 

The theoretical H$\alpha$-to-H$\beta$ ratio for thermally photoinised nebulae has been used 
for de-reddening of measured emission lines in both subsamples. 
 The application of the same de-reddening algorithm to all objects can be justified as follows.
The parameters which are dependent on the absolute fluxes are analysed in the present study
only for the starburst like objects (where this de-reddening  algorithm is correct).  
 As for the AGNs,  this de-reddening algorithm can introduce an 
appreciable error in the de-reddened line intensities. However the de-reddened line 
intensities in the AGNs serve only for classification of those objects.  
The wavelengths of the N\,{\sc ii}]$\lambda$6584 and H$\alpha$ lines
(as well [O\,{\sc iii}]$\lambda$5007 and H$\beta$) are very similar and,
consequently, the [N\,{\sc ii}]$\lambda$6584/H$\alpha$- and 
[O\,{\sc iii}]$\lambda$5007/H$\beta$-line ratios are not that sensitive to the de-reddening 
algorithm. Therefore, the adopted de-reddening algorithm does not result in 
a significant shift of the position of the object in the  
 [N\,{\sc ii}]$\lambda$6584/H$\alpha$ versus [O\,{\sc iii}]$\lambda$5007/H$\beta$ diagram 
and is not likely to result in misclassification of objects.

The [N\,{\sc ii}]$\lambda$6584/H$\alpha$ versus [O\,{\sc iii}]$\lambda$5007/H$\beta$ diagram
is shown in Fig.~\ref{figure:seagull}. 
The upper panel shows data for individual components.
The open (blue) circles are the blue components and the (red) plus signs are 
the red components of the photoionised objects, while
the open (blue) squares are the blue components and the (red) crosses are 
the red components of the AGNs.
The solid line represents the equation
\begin{equation}
\log (\mbox{\rm [O\,{\sc iii}]$\lambda$5007/H$\beta$}) =
\frac{0.61}{\log (\mbox{\rm [[N\,{\sc ii}]$\lambda$6584/H$\alpha$})-0.05} +1.3,
\label{equation:kauff}   
\end{equation}
which separates objects with H\,{\sc ii} spectra from those 
containing an AGN \citep{kauffmannetal2003}.
The long-dashed separation line is the equation
\begin{equation}
\log (\mbox{\rm [O\,{\sc iii}]$\lambda$5007/H$\beta$}) =
\frac{0.61}{\log (\mbox{\rm [[N\,{\sc ii}]$\lambda$6584/H$\alpha$})-0.47} +1.19,
\label{equation:kewley}   
\end{equation}
from \citet{kewleyetal2001}.
The lower  panel in Fig.~\ref{figure:seagull} shows the global data. 
The gray (in the printed version -- light-blue in the colour version) 
filled circles show a large sample of SDSS emission-line  
galaxies from \citet{thuanetal2010}. 
The dark (black) filled circles show our sample of SDSS galaxies with 
double-peaked emission lines (global spectra, sum of blue and red components).
The lines are the same as in the upper panel.
We choose here to use the separation line from \citet{kauffmannetal2003} (see below, Section 4). 

Fig.~\ref{figure:seagull} shows that in a majority cases either both components 
belong to starburst-like objects or to double-peaked AGNs. There seems to be a
strong selection effect. 
Indeed the requirement that a reliable double-Gaussian fit can be obtained 
for each of the emission lines is satisfied for spectra with SDSS resolution only in cases when the
contributions of the blue and red components are similar. 
As a result, our sample contains galaxies with similar 
[N\,{\sc ii}]$\lambda$6584/H$\alpha$- and [O\,{\sc iii}]$\lambda$5007/H$\beta$-fluxes 
for both components. Consequently, both components lie close to each other in 
the [N\,{\sc ii}]$\lambda$6584/H$\alpha$ versus [O\,{\sc iii}]$\lambda$5007/H$\beta$ diagram.

\begin{figure}
\resizebox{1.00\hsize}{!}{\includegraphics[angle=000]{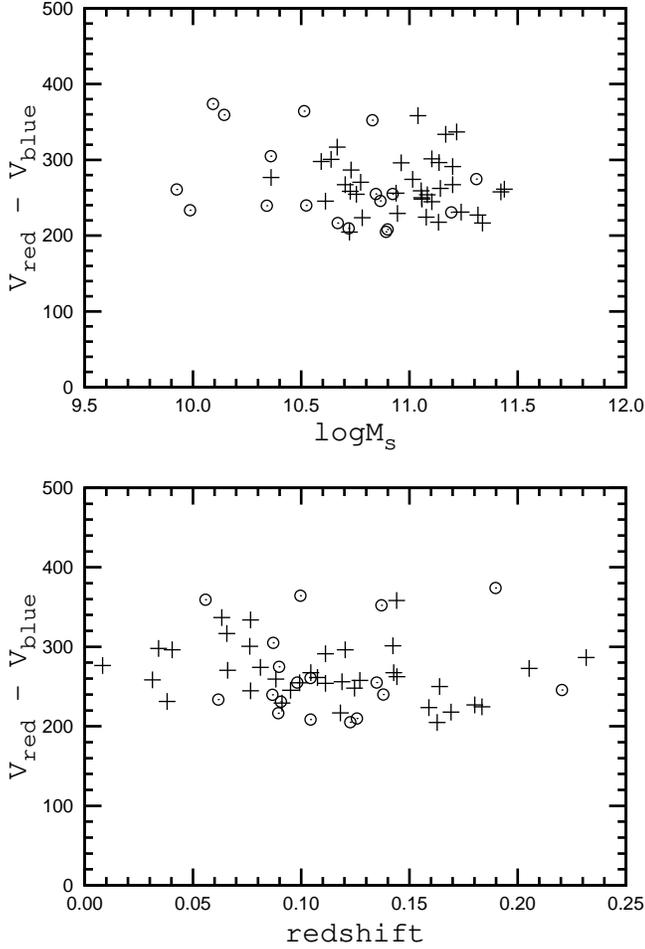}}
\caption{
The difference between radial velocities of blue $V_{\rm \rm blue}$ and red $V_{\rm \rm red}$  
components (in units of km s$^{-1}$) 
as a function of stellar mass of a galaxy ({\it upper panel}) and of 
redshift ({\it lower panel}).
The open circles show galaxies from Sample A.
The plus signs show galaxies from Sample B. 
}
\label{figure:dv}
\end{figure}

\subsection{Velocity separation}

Fig.~\ref{figure:dv} shows the difference between radial velocities of the blue and red 
components as a function of stellar mass ({\it upper panel}) and of 
redshift ({\it lower panel}).The open circles show galaxies from Sample A, while
the plus signs show galaxies from Sample B.
The difference between the radial velocities of the blue and red components is 
determined from the mean value of the differences for individual lines.
The scatter is typically a few per cent, but less than 10\% in all the cases.
Fig.~\ref{figure:dv} shows that  this difference between the radial velocities 
ranges between 200 km s$^{-1}$ and 400 km s$^{-1}$, and that it does not 
correlate with stellar mass, nor with redshift.
However, the upper panel of Fig.~\ref{figure:dv} shows that  
the galaxies with AGN-like spectra are more massive, on average, than
the galaxies with starburst-like spectra. 
 This agrees with the conclusion by \cite{kauffmannetal2003} that AGNs reside 
almost exclusively in massive galaxies. 

\begin{figure}
\resizebox{1.00\hsize}{!}{\includegraphics[angle=000]{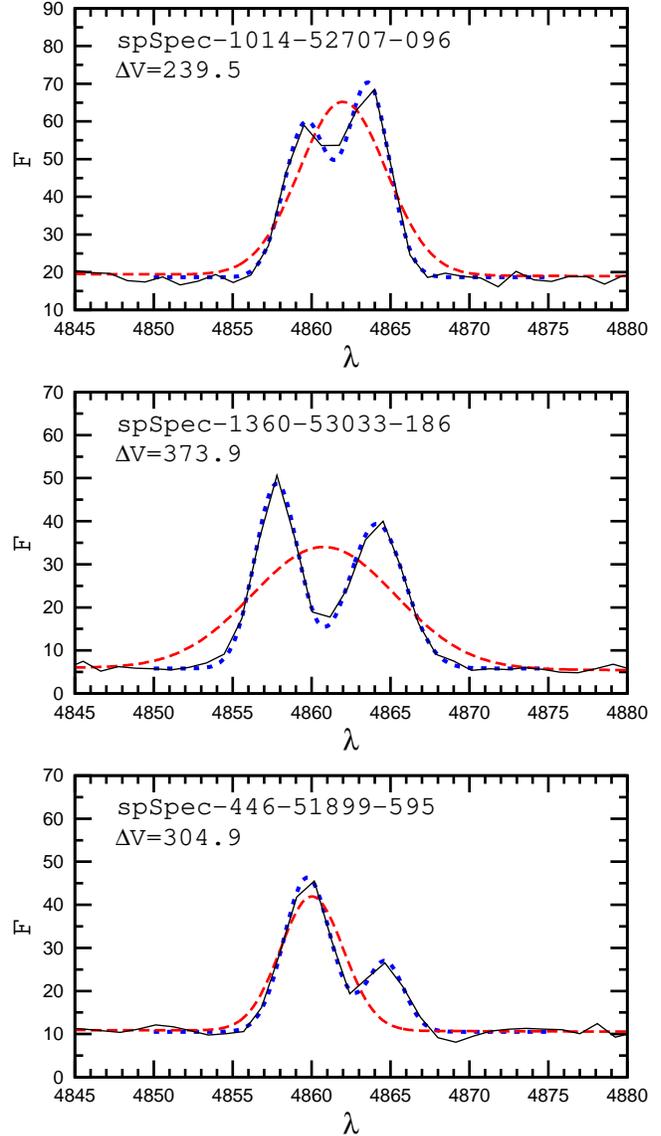}}
\caption{
The approximation of the observed H$\beta$ line profile (dark (black) curve) 
by the single (long-dashed (red) curve) and two (short-dashed (blue) curve) 
Gaussians in spectra of three galaxies from Sample A.
(A color version of this figure is available in the online version.)
}
\label{figure:single}
\end{figure}

\begin{figure}
\resizebox{1.00\hsize}{!}{\includegraphics[angle=000]{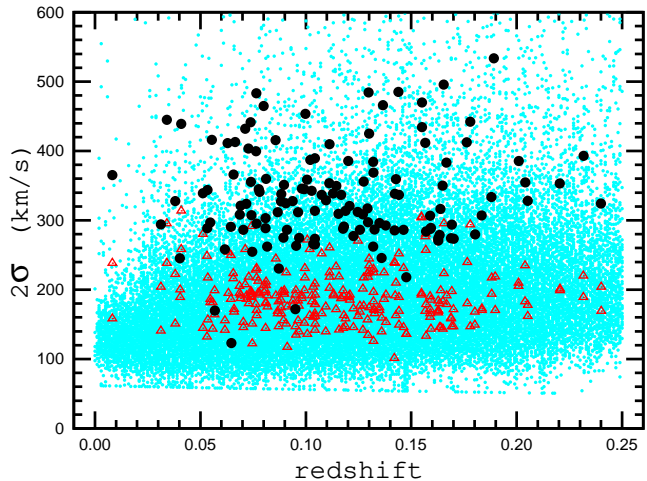}}
\caption{
The Gaussian width parameter $\sigma_{\rm {\rm H}\beta}$ (in units of km/s) as 
a function of redshift for emission-line galaxies taken from the SDSS (gray (blue) filled circles),
for global spectra of  the selected 129 SDSS galaxies with double-peaked emission lines 
(dark (black) filled circles), 
and for the blue and red components of those galaxies (open (red) triangles). 
A color version of this figure is available in the online version.
}
\label{figure:z-vs}
\end{figure}

 Here we will discuss the velocity separations of the double peaks and  
the Gaussian width parameters of the H$\beta$ lines.
It was noted above that the only lines that cannot be reliably 
decomposed into two components are [S\,{\sc ii}]$\lambda$6717 or/and [S\,{\sc ii}]$\lambda$6731 while 
the other lines can be fitted rather well by two Gaussians in 129 selected galaxies. 
Hence, all 129 galaxies will be included in this particular discussion. 
 The lower-limit difference between the radial velocities of the blue $V_{\rm \rm blue}$ and red $V_{\rm \rm red}$  
components ($\sim$200 km s$^{-1}$) in our sample of galaxies is defined by the requirement
that the two peaks in the line profile should be clearly separated. 
\cite{smith2010} have selected active galactic nuclei from the SDSS having double-peaked 
profiles of [O\,{\sc ii}]$\lambda$4959,5007 and other narrow emission lines by visual 
inspection of the quasar spectra. 
The lower-limit velocity separation of the double peaks in their sample is also 
around 200 km s$^{-1}$, i.e. similar to that in our sample of galaxies.

 The rotation velocity of large spiral galaxies is usually above  
200 km s$^{-1}$ and can reach about 300 km s$^{-1}$ (see the compilation by \citet{pilyugin2004}). 
The difference between radial velocities of two H\,{\sc ii} regions in spiral 
galaxy with large inclination can thus be around 400 -- 600 km s$^{-1}$. 
As an alternative interpretation, the large difference between radial velocities of blue 
and red components can be parts of two merging systems.
The largest difference between radial velocities of the blue and red 
components ($\sim$400 km s$^{-1}$) in our sample of galaxies is well within the range given above. 
One may expect that in case many individual H\,{\sc ii} regions are distributed over 
the disc of a giant spiral galaxy with a large inclination, the emission line profile in 
the global single-peaked SDSS spectra of such composite nebulae can have 
a similar width. 

\subsection{Gaussian line width}

We have extracted the line width of the H$\beta$ line ($\sigma_{{\rm H}\beta}$) for 
255,539 emission-line galaxies (with equivalent widths of the H$\beta$ and H$\alpha$ lines 
EW(H$\beta$) $>$ 1.5, EW(H$\alpha$) $>$ 1.5, and 2$\sigma_{{\rm H}\beta}$ $<$  600 km s$^{-1}$) from the 
SDSS database. It should be noted that a single-Gaussian fit to the 
double-peaked emission line can be bad even as a first-order approximation, Fig.~\ref{figure:single}.

Fig.~\ref{figure:z-vs} shows the Gaussian width parameter $\sigma_{{\rm H}\beta}$ (in units of km/s) as 
a function of redshift for a large sample of emission-line galaxies extracted from the SDSS
(gray (blue) filled circles),
for the global spectra of the selected 129 SDSS galaxies with double-peaked emission lines 
(dark (black) filled circles), 
and for the blue and red components of those galaxies (open (red) triangles). 
The values of the Gaussian width parameter $\sigma_{{\rm H}\beta}$ for the global spectra 
(including our sample of 129 galaxies) were taken 
from the SDSS database and measured for the blue and red components. 
Fig.~\ref{figure:z-vs} shows that the line widths of the majority of the 
blue and red components in our sample of galaxies are within range from $\sim$140 km s$^{-1}$ to  
$\sim$220 km s$^{-1}$ and they occupy the same area as the majority of emission-line 
galaxies from the SDSS. However several components have larger widths, up to  $\sim$300 km s$^{-1}$. 
This suggests that some components can themselves be composite nebulae. 
The Gaussian width parameters of the global spectra of our sample of galaxies 
are situated above and along the upper envelope defined by 
the emission-line galaxies from the SDSS. 
However, some of them are situated in the same region as  the majority of the 
blue and red components. These are galaxies where the intensities of 
the blue and red components differ significantly. 
In such case the single-Gaussian fit to the double-peaked emission line reproduces 
the strongest component alone as can be seen in the lower panel of Fig.~\ref{figure:single}.
Thus, the large Gaussian width parameters of the global spectra cannot be a reliable 
criterion for selecting candidates for galaxies with double-peaked emission lines. 
Using this criterion, one will simply miss the galaxies where the intensities of 
the blue and red components differ significantly.

\begin{figure}
\resizebox{1.00\hsize}{!}{\includegraphics[angle=000]{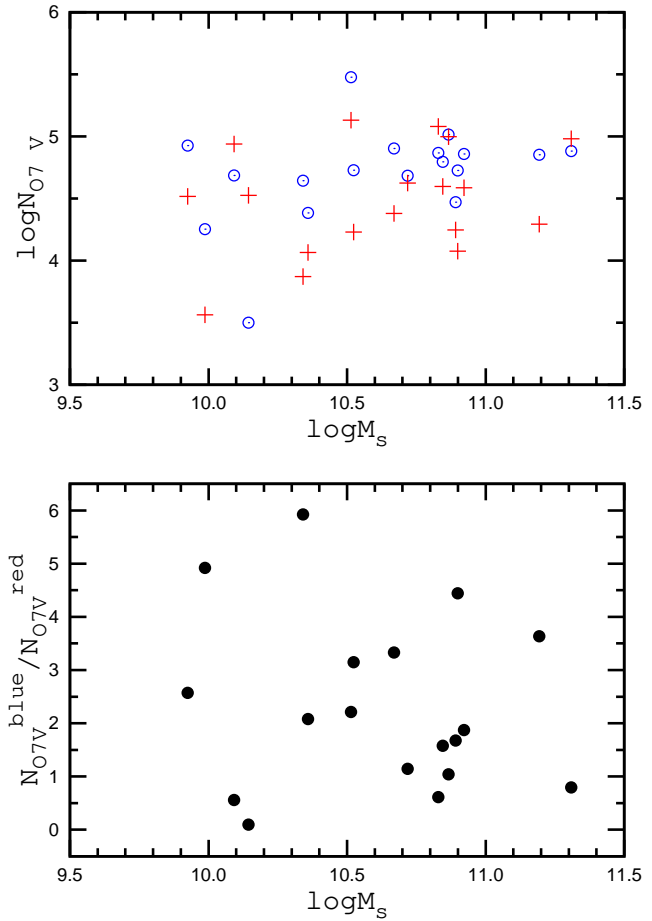}}
\caption{
{\it Upper panel.} The equivalent number of O7V stars  $N_{\rm \rm O7V}$ 
responsible for the excitation 
of H\,{\sc ii} regions versus the host galaxy stellar mass for Sample A. 
The open (blue) circles show the blue components and the (red) plus signs show the 
red components.
{\it Lower panel.} The ratio of equivalent numbers of O7V stars 
responsible for the excitation of the blue and red components.
A color version of this figure is available in the online version.
}
\label{figure:nstar}
\end{figure}

\begin{figure}
\resizebox{1.00\hsize}{!}{\includegraphics[angle=000]{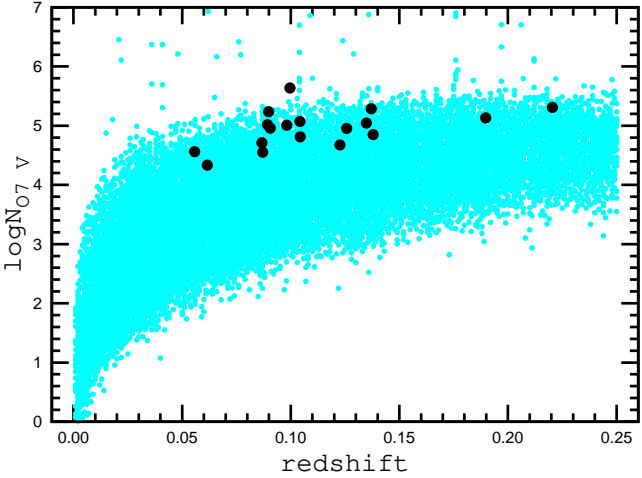}}
\caption{
 The equivalent number of O7V stars  $N_{\rm \rm O7V}$ responsible for the excitation 
of the H\,{\sc ii} regions versus redshift for Sample A (dark (black) filled circles) 
and for star-forming galaxies taken from the SDSS (gray (blue) filled circles). 
A color version of this figure is available in the online version.
}
\label{figure:z-ns}
\end{figure}

\subsection{Starburst strength}

An important characteristic of a H\,{\sc ii} region is the number of the ionising stars it contains. 
Under the assumption of an ionisation-bounded and dust-free nebula, the
H$\beta$ luminosity provides an estimate of the ionising flux. 
The ionising flux can be expressed in terms of the number of so-called 
equivalent O stars of a given subtype responsible for producing the ionising 
luminosity. The number of zero-age main sequence O7V stars, N$_{\rm \rm O7V}$, is 
usually used to specify the ionising flux. The value of N$_{\rm \rm O7V}$ can be 
easily derived from the observed H$\beta$ luminosity and the Lyman continuum 
flux of an individual O7V star. The number of ionising photons from an O7V  star is   
$N_{\rm \rm Lc}$ = 5.62$\times$10$^{48}$ s$^{-1}$ \citep{martinsetal2005}
or $N_{\rm \rm Lc}$ = 1.12$\times$10$^{49}$ s$^{-1}$ \citep{vacca1994}. 
Here the value of $N_{\rm \rm Lc}$ after \citet{martinsetal2005} has been adopted. 
If we instead adopt the value of $N_{\rm \rm Lc}$ after \citet{vacca1994} then  
the number of equivalent O stars are increased by about a factor two.   
One ionising photon from the star produces 0.157 H$\beta$ photons (0.449 H$\alpha$ 
photons) from the  H\,{\sc ii} region \citep{osterbrock2006}.    
The density-bounded dust-free H\,{\sc ii} region excited by one  O7V star will 
have the H$\beta$ luminosity 3.38$\times$10$^{36}$ erg s$^{-1}$  or log(L$_{\rm {\rm H}\beta}$/L$_{\rm \sun}$) = 2.93 and 
the H$\alpha$ luminosity 7.64$\times$10$^{36}$ erg  s$^{-1}$  or log(L$_{\rm {\rm H}\alpha}$/L$_{\rm \sun}$) = 3.30. 
When these conditions (density-bounded and/or dust-free) are not met, the values of the 
number of equivalent O stars obtained here should be interpreted as lower limits.   

The reddening-corrected  H$\beta$ flux $F_{\rm {\rm H}\beta}^O$ is obtained from 
the observed  $F_{\rm {\rm H}\beta}^{obs}$ flux using the relation  \citep{blagrave2007} 
\begin{equation}
\log(F_{\rm {\rm H}\beta}^{obs}/F_{\rm {\rm H}\beta}^O) = - C_{\rm \rm H\beta} ,  
\label{equation:hbcorr}   
\end{equation}
where  C(H$\beta$) is  the extinction coefficient.

The distances to SDSS galaxies are calculated from 
\begin{equation}
d = \frac{cz}{H_0} ,
\label{equation:d}   
\end{equation}
where $d$ is the distance in Mpc,  $c$ the speed of light in km s$^{-1}$, and $z$ the redshift. 
$H_0$ is the Hubble constant, here assumed to be equal to  72 ($\pm$8) km s$^{-1}$ Mpc$^{-1}$ 
\citep{freedman2001}. 
Since objects from Sample A have low redshifts (only two objects have $z$ $>$ 0.15) 
the low-redshift approximation for distance determination has been used. 

The upper panel of Fig.~\ref{figure:nstar} shows the equivalent number of O7V stars  
$N_{\rm \rm O7V}$ responsible for the excitation of blue and red components in galaxies 
from Sample A. The lower panel shows the ratio of equivalent numbers of O7V stars  
responsible for the excitation of blue and red components.
Inspection of  Fig.~\ref{figure:nstar} shows the typical value of the equivalent number 
of O7V stars in starbursts in galaxies from Sample A is in the range 10$^4$ -- 10$^5$.
\citet{kennicutt1988} has given the mean H$\alpha$ fluxes for the three 
brightest H\,{\sc ii} regions in a sample of nearby galaxies. 
Those fluxes, converted to equivalent numbers of O7V stars, correspond to 
N$_{\rm \rm O7V}$ $\sim 10^2$. Thus, both the blue and red components in our sample of SDSS 
galaxies contain many more (up to 2-3 orders of magnitude) 
ionising stars, when compared to the brightest H\,{\sc ii} regions in nearby galaxies. 
Fig.~\ref{figure:z-ns} shows a comparison of the equivalent number of O7V stars  
$N_{\rm \rm O7V}$ for Sample A (the sum of the $N_{\rm \rm O7V}$ for blue and red components) 
and for a large sample of star-forming galaxies from SDSS. This sample of galaxies were 
selected from the MPA/JHU catalogs using  the dividing line between starburst like objects 
and AGNs from \citet{kauffmannetal2003}.
The reddening-corrected  H$\beta$ flux is obtained using  Eq.(\ref{equation:hbcorr}). 
Only the galaxies with EW(H$\beta$) $>$ 1.5 and EW(H$\alpha$) $>$ 1.5 are shown. 
 Fig.~\ref{figure:z-ns} also shows 
that the number of ionizing stars in the galaxies with two starbursts from our Sample A are 
similar to that of galaxies (from the SDSS) with a large amount of ionising stars.

\begin{figure}
\resizebox{1.00\hsize}{!}{\includegraphics[angle=000]{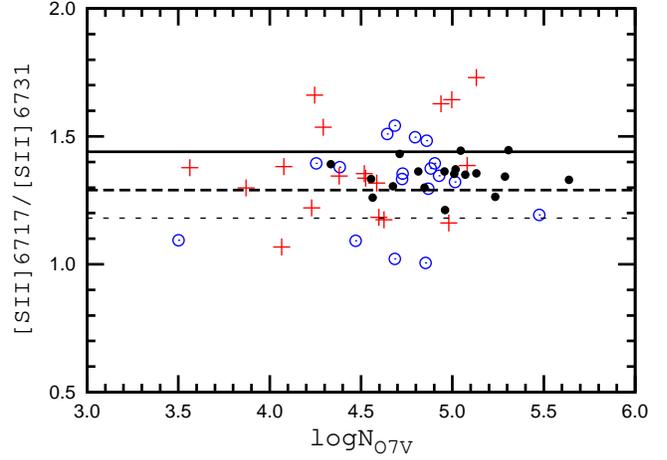}}
\caption{
The density-sensitive [S\,{\sc ii}]$\lambda$6717/[S\,{\sc ii}]$\lambda$6731 line ratio versus 
the equivalent number of O7V stars  $N_{\rm \rm O7V}$ responsible for the excitation 
of H\,{\sc ii} regions. 
The open (blue) circles show the blue components. 
The (red) plus signs show the red components. 
The filled circles show the sum of blue and red components. 
The solid line shows the zero-density limit ($N_e =1$~cm$^{-3}$), 
the dashed line corresponds to the electron density $N_e =100$~cm$^{-3}$, 
and the dotted line corresponds to the electron density $N_e = 200$~cm$^{-3}$.
A color version of this figure is available in the online version.
}
\label{figure:ne}
\end{figure}

\subsection{Electron density}

Fig.~\ref{figure:ne} shows the density-sensitive  
[S\,{\sc ii}]$\lambda$6717/[S\,{\sc ii}]$\lambda$6731 line ratio 
as a function of the equivalent number of O7V stars  $N_{\rm \rm O7V}$ in the starbursts. 
The open (blue) circles show the blue components. The (red) plus signs show the red components. 
The filled circles show the data from global spectra (sum of blue and red components). 
The expected zero density limit ([S\,{\sc ii}]$\lambda$6717/[S\,{\sc ii}]$\lambda$6731 = 1.44 
at $N_e =1$~cm$^{-3}$ with $t_2 = 1.0$), is shown by the solid line. 
The dashed and dotted lines show the line ratios corresponding to   
$N_e =100$~cm$^{-3}$ ([S\,{\sc ii}]$\lambda$6717/[S\,{\sc ii}]$\lambda$6731 = 1.29) and 
$N_e =200$~cm$^{-3}$ ([S\,{\sc ii}]$\lambda$6717/[S\,{\sc ii}]$\lambda$6731 = 1.18), respectively.

The density-sensitive [S\,{\sc ii}]$\lambda$6717/[S\,{\sc ii}]$\lambda$6731 line ratio 
in the global spectra of all galaxies in our Sample A   
show an electron density $N_e \leq 200$~cm$^{-3}$ 
(filled circles in Fig.~\ref{figure:ne}), with   
the majority of them having $N_e \leq 100$~cm$^{-3}$. 
Hence, they are all in the low-density regime, as 
is typical for the majority of extragalactic H\,{\sc ii} regions 
\citep{zkh,bresolinetal05,gutierrez2010}. 
The  [S\,{\sc ii}]$\lambda$6717/[S\,{\sc ii}]$\lambda$6731 line ratio 
in the blue and red components show a larger scatter (see Fig.~\ref{figure:ne}), which
may be caused by the uncertainties in the line decomposition.

\begin{figure}
\resizebox{1.00\hsize}{!}{\includegraphics[angle=000]{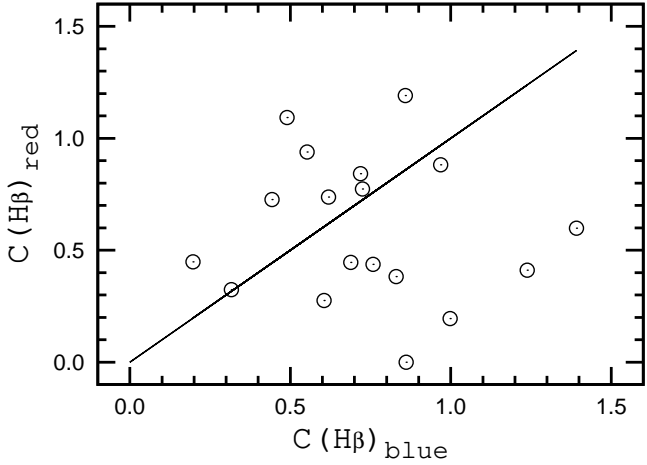}}
\caption{
The extinction C(H$\beta$) in the blue component versus the extinction C(H$\beta$) 
in the red component for galaxies from Sample A.
}
\label{figure:chb}
\end{figure}

Two giant  H\,{\sc ii} regions located at different positions inside the disc (one 
 H\,{\sc ii} regions can be associated with the circumnuclear star formation) may 
be responsible for the double-peaked emission lines in Sample A.  
However, the double-peaked emission lines in global spectra can be caused not only 
by two starbursts in the same galaxy, but also two starbursts in two different 
galaxies, provided these galaxies are very closely located on the sky (projected on top 
of each other and thus detected within the same SDSS fiber). 
If the radiation from a star-forming region in a more distant galaxy 
passes through a less distant galaxy, one may expect that the extinction of the
red component (star-forming region in the more distant galaxy) should be larger
than the extinction of the blue component (star-forming region in the less distant galaxy). 
Fig.~\ref{figure:chb} shows the extinction coefficient C(H$\beta$) in the blue component versus 
the C(H$\beta$) in the red component for galaxies in Sample A.
Fig.~\ref{figure:chb} also shows that there is no systematic difference 
between the values of the extinction coefficient C(H$\beta$) in the blue and red components.
It should be noted, however, that it cannot be excluded that the more distant 
galaxies can have slightly lower radial velocities than less distant ones due to 
the random component of their radial velocities.

\begin{figure*}
\resizebox{1.00\hsize}{!}{\includegraphics[angle=000]{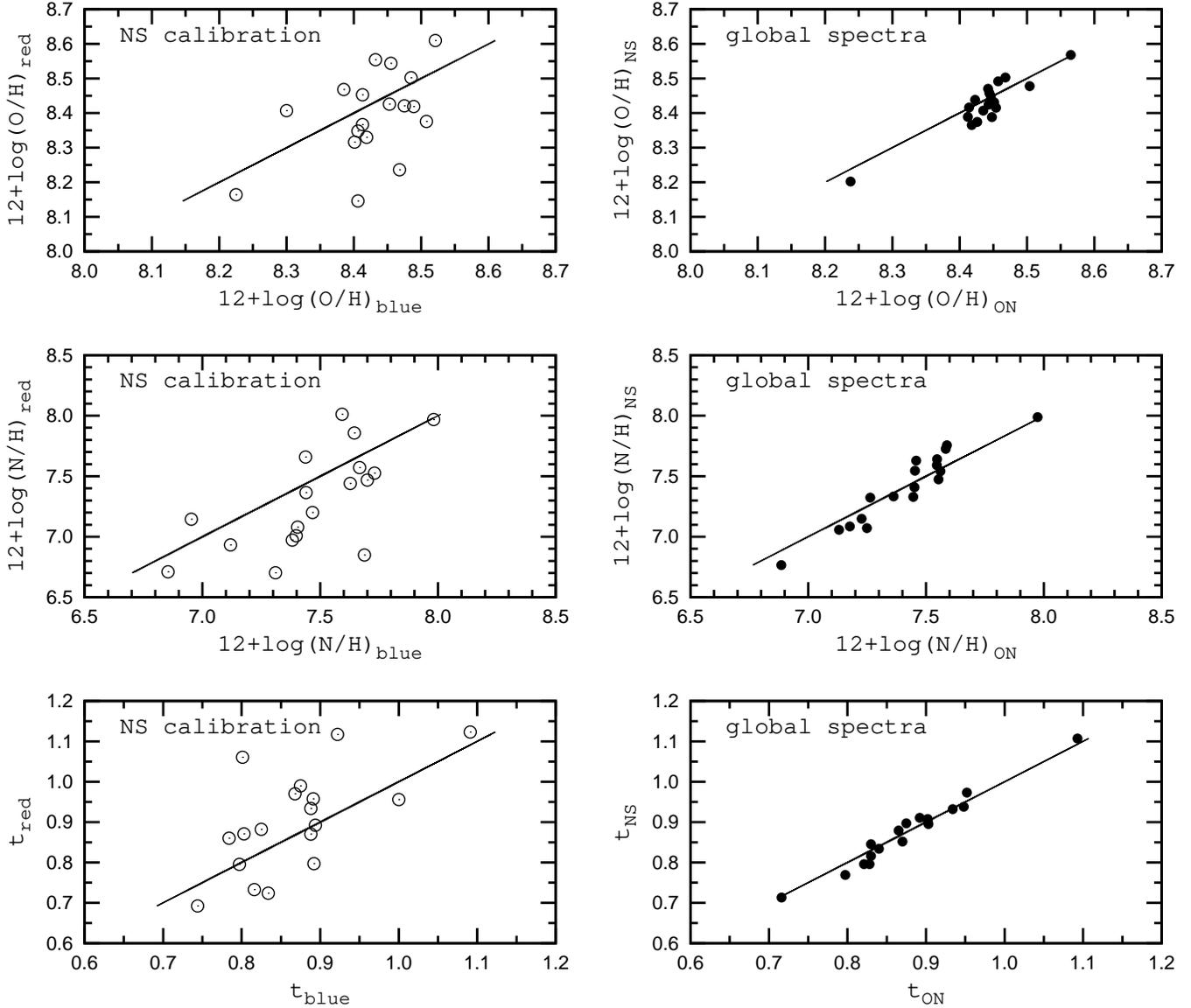}}
\caption{
{\it Left column panels.}
Comparison of oxygen abundances (upper panel), nitrogen abundances 
(middle panel) and $t_2$ electron temperatures  (lower panel) in blue components  
and the same in the red components of the galaxies from Sample A. 
{\it Right column panels.}
Comparison oxygen abundances (upper panel), nitrogen abundances 
(middle panel) and $t_2$ electron temperatures  (lower panel) determined from global 
line fluxes using the NS-calibration with the same derived using the ON-calibration. 
The solid lines represent equal values. 
}
\label{figure:zz}
\end{figure*}

\section{Abundances and temperatures}

It is impossible to divide the [O\,{\sc ii}]$\lambda$3727+$\lambda$3729 doublet 
into a blue and a red component at the SDSS spectral resolution. 
Therefore, we have used NS-calibration to determine 
abundances and the $t_2$ electron temperatures in the blue and red components. 
The NS-calibration is an empirical calibration for determination of oxygen and
nitrogen abundances as well as electron temperatures in H\,{\sc ii} regions where the
nebular oxygen line [O\,{\sc ii}]$\lambda$3727+$\lambda$3729 is not available 
\citep{pilyuginmattsson2011}.
The NS-calibration relations express the abundances (and electron temperatures) 
in terms of the fluxes in the strong emission lines of O$^{++}$,
N$^+$, and S$^+$  and have been derived using spectra of
H\,{\sc ii} regions with well-measured electron temperatures as calibration datapoints. 
The NS-calibration provides reliable oxygen and nitrogen
abundances for H\,{\sc ii} regions of all metallicities.
The resultant NS-calibration abundances and electron temperatures 
are given in Table~\ref{table:list}. 

The left panels of Fig.~\ref{figure:zz} show
the comparison of oxygen abundances (upper panel), nitrogen abundances 
(middle panel) and $t_2$ electron temperatures  (lower panel) in the blue components  
and the same in the red components of galaxies from Sample A. 
Fig.~\ref{figure:zz} also shows that 
the oxygen and nitrogen abundances and $t_2$ electron temperatures  in the blue 
and red components are close to each other, which may reflect a
strong selection effect. Indeed, as it has already been noted above,  
at the limited SDSS spectral resolution, more or less reliable double-Gaussian fits to 
the emission lines  ([\,{\sc iii}]$\lambda$5007, [N\,{\sc ii}]$\lambda$6584, 
[S\,{\sc ii}]$\lambda$6717 and [S\,{\sc ii}]$\lambda$6731) 
can be obtained if the blue and red components make comparable contributions 
to the "global lines". Hence, we have selected spectra with double-peaked 
emission lines where the intensities of all these lines in the blue and red components 
are similar and, consequently, the abundances are similar too.    

Since the SDSS spectra are closer to global spectra of galaxies rather than to 
spectra of individual H\,{\sc ii} regions, the abundances determined 
from the SDSS spectra are "global abundances". For our sample of galaxies we can 
determine the "global flux" in every emission line as the sum of fluxes 
from the blue and red components, and estimate the global abundances using
the NS-calibration.
However, as the global flux in the [O\,{\sc ii}]$\lambda$3727+$\lambda$3729 doublet can be 
measured in the spectra considered here, it is possible to estimate the global oxygen and nitrogen abundances 
and $t_2$ electron temperatures  using another empirical calibration, 
the ON-calibration \citep{pilyuginetal2010}. 
The ON-calibration relations express the abundances (and electron temperatures) 
in terms of the fluxes in the strong emission lines O$^{++}$, O$^{+}$, and 
N$^+$. It has also been derived using the spectra of
H\,{\sc ii} regions with well-measured electron temperatures as calibration datapoints. 

The resultant global abundances and electron temperatures are given in Table~\ref{table:list}. 
The right panels of Fig.~\ref{figure:zz} show
a comparison of global oxygen abundances (upper panel), nitrogen abundances 
(middle panel) and $t_2$ electron temperatures  (lower panel) in Sample A
determined using the NS- and ON-calibrations.  Fig.~\ref{figure:zz} also shows that 
the oxygen and nitrogen abundances, as well as the $t_2$ electron temperatures,  
derived using the NS- and ON-calibrations are in good agreement, which can be considered  
as evidence that our abundances and temperatures are realistic.

\begin{figure}
\resizebox{1.00\hsize}{!}{\includegraphics[angle=000]{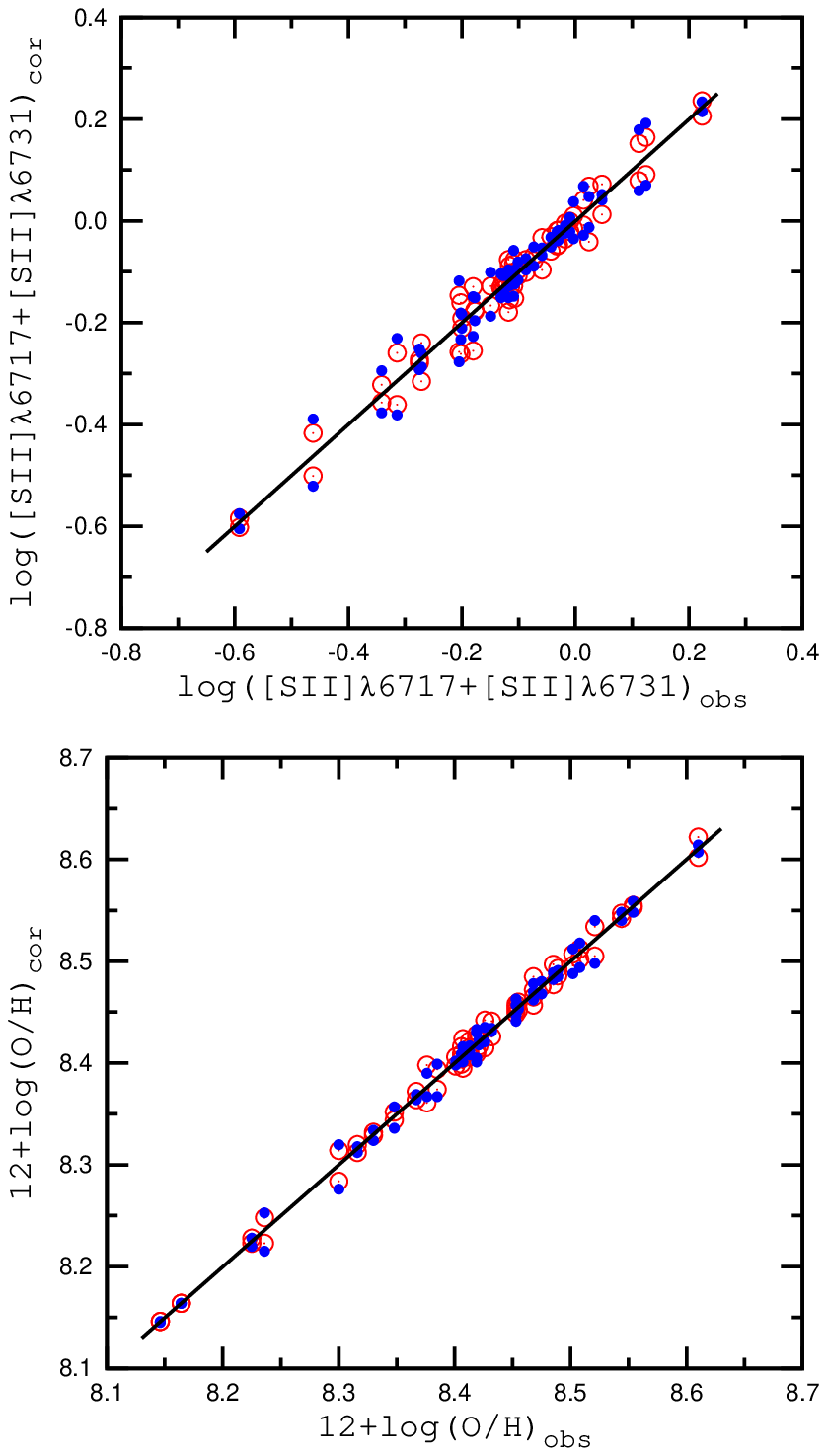}}
\caption{ 
{\it Upper panel.}
The corrected flux ($S_2$)$_{\rm cor}$ against the observed flux ($S_2$)$_{\rm obs}$ 
in the sulfur lines for the blue and red components in the galaxies in Sample A. 
The open (red) circles show the case when the corrected values have been estimated assuming  
the "true" line ratio [S\,{\sc ii}]$\lambda$6717/[S\,{\sc ii}]$\lambda$6731 = 1.44 
which corresponds to the electron density $N_e = 100$~cm$^{-3}$ (see text).  
The filled (blue) circles show the corrected values obtained assuming  
the "true" line ratio [S\,{\sc ii}]$\lambda$6717/[S\,{\sc ii}]$\lambda$6731 = 1.29 
which corresponds to the electron density $N_e = 1$~cm$^{-3}$. 
The solid line shows the case of equal values. 
{\it Lower panel.} 
The corrected oxygen abundances (O/H)$_{\rm NS, cor}$  against the observed abundnaces (O/H)$_{\rm NS, obs}$  
in the blue and red components  in the galaxies in Sample A.
The meaning of the symbols are the same as in the upper panel.
The solid line shows the case of equal values.
(A color version of this figure is available in the online version.)
}
\label{figure:ohoh}
\end{figure}

As it was noted above, the  [S\,{\sc ii}]$\lambda$6717/[S\,{\sc ii}]$\lambda$6731 
line ratio in the blue and red components show a larger scatter than that in the global 
spectra (see Fig.~\ref{figure:ne}), which may be caused by the uncertainties in the line 
decomposition. How do such uncertainties affect the derived abundances in the blue and 
red components? We argue it can be estimated in the following way. 
It is known that the majority of extragalactic H\,{\sc ii} regions are in the low-density 
regime \citep{zkh,bresolinetal05,gutierrez2010}.  The global spectra of all galaxies in 
our Sample A also show a low electron density having $N_e \leq 100$~cm$^{-3}$ (Fig.~\ref{figure:ne}). 
Then one can expect that the blue and red components have also a  low electron density 
and, consequently, the $Rs$ = [S\,{\sc ii}]$\lambda$6717/[S\,{\sc ii}]$\lambda$6731 line ratio 
in their spectra should be within the interval defined by the  $Rs$ values at 
 $N_e$ = 100~cm$^{-3}$ and $N_e$ = 1~cm$^{-3}$.

Let us first assume $Rs$ at $N_e$ = 100~cm$^{-3}$ is the``true''  value ($Rs_{\rm true}$). 
Then the uncertainties in the line ratio in the blue components 
can be quantified by the value $c_b$ = $Rs_{\rm b}$/$Rs_{\rm true}$. If the $c_b$ is larger/smaller than unity then 
the flux in the [S\,{\sc ii}]$\lambda$6717 line is over-/underestimated by a factor of $c_{\rm b,1}$  
up to  $c_b$ and/or the flux in the [S\,{\sc ii}]$\lambda$6731 line is under-/overestimated by a 
factor of $c_{\rm b,2}$  up to  $c_b$ so that  $c_{\rm b,1}$/$c_{\rm b,2}$ = $c_{\rm b}$. One can estimate 
the corrected flux in the sulfur lines as
($S_2$)$_{\rm cor}$ = [S\,{\sc ii}]$\lambda$6717/$c_{\rm b,1}$ + [S\,{\sc ii}]$\lambda$6731/$c_{\rm b,2}$. 
It is evident that the error in the $S_2$ is maximised when $c_{\rm b,1}$ = $c_{\rm b}$ and $c_{\rm b,2}$ = 1 
if [S\,{\sc ii}]$\lambda$6717 $>$ [S\,{\sc ii}]$\lambda$6731 and 
 when $c_{\rm b,1}$ = 1 and $c_{\rm b,2}$ =  1/$c_{\rm b}$  
if [S\,{\sc ii}]$\lambda$6717 $<$ [S\,{\sc ii}]$\lambda$6731. 
We have then considered the two exreme cases. We have found two values of the corrected flux 
in the sulfur lines.  
The uncertainties in this line ratio in the red components 
can be quantified by the value $c_r$ = $Rs_{\rm r}$/$Rs_{\rm true}$. 
The  corrected flux ($S_2$)$_{\rm cor}$  in the sulfur lines can be estimated in the same way for 
the red components.
The obtained values of $c_{\rm b}$ and $c_{\rm r}$ for galaxies in Sample A are all in the range
from $\sim$0.7 to $\sim$1.3. As a result, the difference between the corrected flux ($S_2$)$_{\rm cor}$ 
and the observed flux ($S_2$)$_{\rm obs}$ in the sulfur lines 
is not exceeding $\sim$15\% for any object.  
We plot the corrected flux ($S_2$)$_{\rm cor}$  against the observed one ($S_2$)$_{\rm obs}$  in 
the upper panel of  Fig.~\ref{figure:ohoh} (open (red) circles).

Let us now assume that the $Rs$ at $N_e$ = 1~cm$^{-3}$ is the  ``true'' $Rs_{\rm true}$  value. 
We plot the corrected flux ($S_2$)$_{\rm cor}$  against the observed one ($S_2$)$_{\rm obs}$  in 
the upper panel of  Fig.~\ref{figure:ohoh} (filled (blue) circles). 
The obtained values of $c_{\rm b}$ and $c_{\rm r}$ for galaxies in Sample A are again within the range
from $\sim$0.7 to $\sim$1.3 and the difference between the corrected flux ($S_2$)$_{\rm cor}$ 
and the observed flux ($S_2$)$_{\rm obs}$ in the sulfur lines for blue and red components 
is never exceeding $\sim$15\%.  

One may consider the measured global $Rs_{\rm g}$ = [S\,{\sc ii}]$\lambda$6717/[S\,{\sc ii}]$\lambda$6731 
as being the ``true'' value instead of the $Rs$ at $N_e$ = 100~cm$^{-3}$ and at $N_e$ = 1~cm$^{-3}$. 
In such case the differences between the corrected fluxes ($S_2$)$_{\rm cor}$ 
and flux ($S_2$)$_{\rm obs}$ in the sulfur lines in the blue and red components 
are usually less than $\sim$10\%.   

We have found the corrected oxygen abundances (O/H)$_{\rm NS, cor}$ in blue and 
red components through the NS-calibration using the values of corrected  
flux ($S_2$)$_{\rm cor}$  in the sulfur lines. The corrected oxygen abundances (O/H)$_{\rm NS, cor}$  
against the observed abundances (O/H)$_{\rm NS, obs}$ are shown in  
the lower panel of Fig.~\ref{figure:ohoh}. 
The lower panel of Fig.~\ref{figure:ohoh} shows that the difference 
between the corrected oxygen abundance (O/H)$_{\rm NS, cor}$ and the observed abundance
(O/H)$_{\rm NS, obs}$ is less than 0.03 dex in all cases.
This is because the coefficients of the terms with log$S_2$ in the NS-calibration 
relations for oxygen abundance determinations are less than unity for all classes of  
H\,{\sc ii} regions (hot, warm and cold) \citep{pilyuginmattsson2011}. 
Consequently, the uncertainty in $S_2$ which is less than $<$ 15\% (or less than $\sim$0.07 dex) 
results in an error in log(O/H)$_{\rm NS, cor}$  less than 0.03 dex.
The difference between the corrected nitrogen abundance (N/H)$_{\rm NS, cor}$ and the observed abundance
(N/H)$_{\rm NS, obs}$ is slightly larger, up tp 0.07 dex.
This is because the coefficients of the terms with log$S_2$ in the NS-calibration 
relations for nitrogen abundance determinations are close to unity for cold and warm 
H\,{\sc ii} regions  \citep{pilyuginmattsson2011}.
Hence, the given value of the uncertainty in $S_2$ (up to 15\% or up to 0.07 dex) results in a similar error in log(N/H)$_{\rm NS, cor}$.  
Thus, the errors in the (O/H)$_{\rm NS}$ and the (N/H)$_{\rm NS}$  abundances caused by the uncertainties in the sulfur line 
decomposition are relatively small and cannot affect the results significantly.

\begin{figure}
\resizebox{1.00\hsize}{!}{\includegraphics[angle=000]{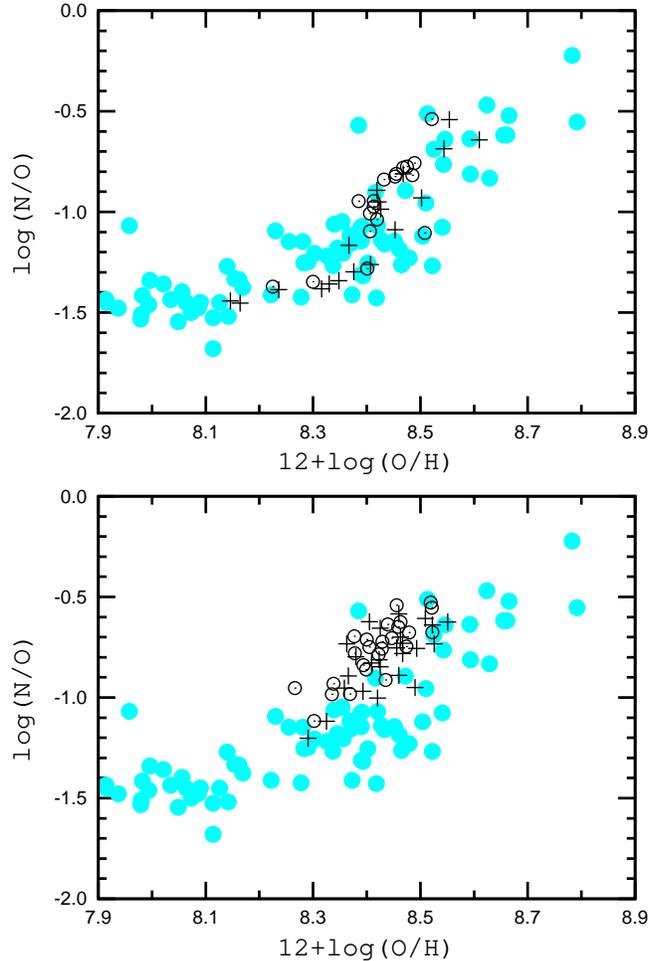}}
\caption{
The O/H--N/O diagram.
{\it Upper panel.}
The open circles show abundances derived using the NS-calibration for blue 
components of galaxies of the Sample A, the plus signs show that 
for red components. The filled gray (light-blue) circles show
$T_e$-based abundances in the sample of best-studied H\,{\sc ii}
regions in nearby galaxies (the compilation of data from \citet{pilyuginetal2010}). 
{\it Lower panel.}
The same as the upper panel but for galaxies which lie in 
the [N\,{\sc ii}]$\lambda$6584/H$\alpha$ versus [O\,{\sc iii}]$\lambda$5007/H$\beta$ diagram
between the separation lines after \citep{kauffmannetal2003} and after 
\citet{kewleyetal2001} (see Fig.~\ref{figure:seagull}). 
(A color version of this figure is available in the online version.)
}
\label{figure:ohno}
\end{figure}

The O/H--N/O diagram provides an additional possibility to test the 
correctness of the determined oxygen and nitrogen abundances. 
The upper panel of Fig.~\ref{figure:ohno} shows the O/H - N/O diagram for Sample A. 
The open circles show abundances derived using the NS-calibration for blue 
components, the plus signs show the same for red components.
The filled gray (light-blue) circles show
$T_e$-based abundances in the calibration sample of H\,{\sc ii}
regions in nearby galaxies (the compilation of data from \citet{pilyuginetal2010}). 
Fig.~\ref{figure:ohno} also shows that the abundances derived using 
the NS-calibration occupy the same region in the  O/H - N/O diagram as the
$T_e$-based abundances of the calibration sample, which implies that our 
NS-calibration abundances are correct.

The exact location of the dividing line between H\,{\sc ii} regions 
and AGNs is still controversial \citep[see, e.g.,][]{kewleyetal2001,kauffmannetal2003,stasinskaetal2006}. 
The objects that lie between the dividing lines according \citep{kauffmannetal2003} and 
\citet{kewleyetal2001}  in the [N\,{\sc ii}]$\lambda$6584/H$\alpha$ versus [O\,{\sc iii}]$\lambda$5007/H$\beta$ 
diagram (see Fig.~\ref{figure:seagull}) are starburst like objects if the dividing line according to 
\citet{kewleyetal2001} is used but they are AGN-like objects 
(or, at least, they are not purely thermally photoionised objects) 
if the dividing line according to \citet{kauffmannetal2003} is used. 
Note, however, that the dividing line suggested by \citet{stasinskaetal2006} is very similar to that 
of \citet{kauffmannetal2003}.

The O/H - N/O diagram may also provide an indirect way to decide which curve 
[that according to \citet{kauffmannetal2003} or that according to \citet{kewleyetal2001}]
outlines best the area occupied by starburst-like objects in the BPT diagram. 
If objects that lie between these lines are starburst-like objects, the emission line 
fluxes in their spectra correspond to thermally photoionised objects.
The NS-calibration (and other strong-line calibrations) developed for thermally 
photoionised nebulae will then provide reliable oxygen and nitrogen abundances 
for these objects, which will occupy the same region in the O/H - N/O diagram 
as H\,{\sc ii} regions in nearby galaxies. 
If, on the other hand,  the [N\,{\sc ii}] fluxes 
are produced by non-thermal radiation or at least are enhanced
by the contribution of non-thermal radiation, using the 
NS-calibration on these nebulae will result in too high nitrogen abundances.   
One may expect that the positions in the O/H -- N/O diagram will be  
shifted significantly as compared to the positions of thermally photoionised  
H\,{\sc ii} regions in nearby galaxies. 
The lower panel of Fig.~\ref{figure:ohno} shows the O/H -- N/O diagram for 
galaxies  which lie between the two dividing lines in the 
[N\,{\sc ii}]$\lambda$6584/H$\alpha$ versus [O\,{\sc iii}]$\lambda$5007/H$\beta$ 
diagram together with the positions of thermally photoionised  
H\,{\sc ii} regions in nearby galaxies. 
The positions of the galaxies considered here show a systematic shift toward higher N/O-ratio
compared to the positions of the thermally photoionised  H\,{\sc ii} regions in nearby galaxies, 
which suggest that the emission lines are distorted by the contribution of  
non-thermal radiation, i.e. they are not purely thermal phoionisation objects. 
If this is the case, the line according to \citep{kauffmannetal2003} is to be favoured 
as it outlines the area occupied by certainly starburst-like objects in the 
BPT diagram.
It has been assumed \cite[e.g., by][]{juneau2011} that the line from  \citet{kewleyetal2001}
can be considered as the curve outlining the area occupied by ``pure'' AGNs in the 
BPT diagram, and that objects which lie between the curves according to \citet{kauffmannetal2003} 
and \citet{kewleyetal2001} are composite objects 
where both a starburst and an AGN make contributions. However, this interpretation is 
not indisputable \citep{cidfernandes2010,cidfernandes2011}.

\begin{figure}
\resizebox{1.00\hsize}{!}{\includegraphics[angle=000]{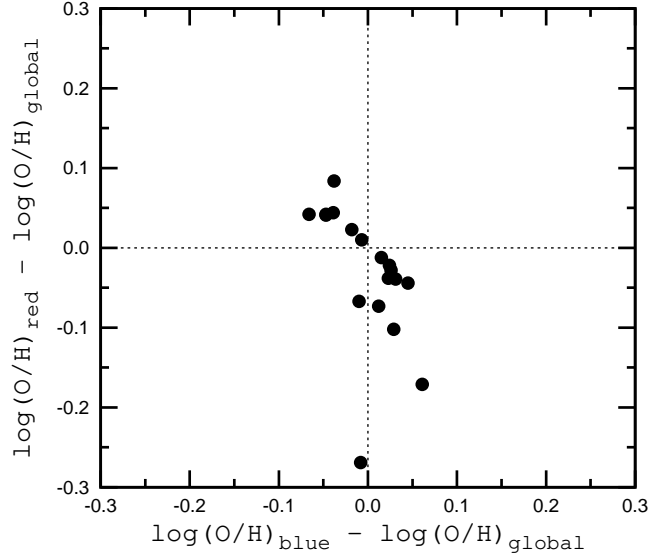}}
\caption{
The difference between oxygen abundances in the blue component and the global 
oxygen abundances versus 
the difference between oxygen abundances in the red component and the global 
oxygen abundances for galaxies in Sample A.
}
\label{figure:doh}
\end{figure}

Comparison of oxygen abundances in the blue and red components and the
global oxygen abundances tell us something about how representative/correct
the abundances derived from the global spectra are. 
Fig.~\ref{figure:doh} shows the difference between oxygen abundances in the blue 
component and the global oxygen abundances versus 
the difference between oxygen abundances in the red component and the global 
oxygen abundances for galaxies in Sample A.
Furthermore, Fig.~\ref{figure:doh} clearly shows that there is an
anti-correlation between the two abundance differences described above. This means the 
global oxygen abundance obtained using the NS-calibration is typically in between   
the oxygen abundance of the blue and red components.

\section{Conclusions}

We have extracted several hundred galaxies from the SDSS spectral database
with double-peaked emission lines in their global spectra. Among other possibilities (e.g. 
double-peaked AGNs), one 
may expect such line profiles if two strong starbursts take place simultaneously in 
a galaxy. We have fitted the emission lines (H${\alpha}$, H${\beta}$, [O\,{\sc iii}]$\lambda$5007, 
[N\,{\sc ii}]$\lambda$6584, [S\,{\sc ii}]$\lambda$6717 and [S\,{\sc ii}]$\lambda$6731) by two Gaussians
in 129 spectra to separate the flux of the two (blue and red) components. A more or less reliable 
decomposition of the emission lines has been found for 55 spectra. 

Using the standard classification  
[N\,{\sc ii}]$\lambda$6584/H$\alpha$ versus [O\,{\sc iii}]$\lambda$5007/H$\beta$ 
diagram and the dividing lines after \citet{kauffmannetal2003}, 
we have divided the galaxies from our sample 
into two subsamples: a Sample A consisting  of 18 galaxies  where both components  
belong to the photoionised objects and a Sample B containing 37 galaxies 
which show nonthermal ionisation (AGN). 
All blue and red components have narrow lines regardless of their positions in the 
[N\,{\sc ii}]$\lambda$6584/H$\alpha$ versus [O\,{\sc iii}]$\lambda$5007/H$\beta$ diagram. 

The differences between radial velocities of the blue and red components lie between 200 and 400 
km s$^{-1}$ for both subsamples. The equivalent number of ionising stars is 
within the range 10$^4$ -- 10$^5$ O7V stars for each component in Sample A.
 
We have estimated the oxygen and nitrogen abundances as well as the electron 
temperatures for each component, and for the global spectra in Sample A, using the recent NS-calibration.
We have found that  the global oxygen abundance is typically in between the oxygen abundance of 
the blue and red components.
This conclusion is based on a small sample of galaxies
which, by selection, have red and blue line components of similar flux. 
To confirm (or reject) our conclusion a larger sample of galaxies 
should be considered.    
Applying the ON-calibration on the global spectra shows that the ON-calibration and NS-calibration
give oxygen and nitrogen abundances and electron temperatures which are in good agreement. 

The O/H - N/O diagram for the galaxies in our sample gives indirect evidence suggesting that the dividing 
line in the [N\,{\sc ii}]$\lambda$6584/H$\alpha$ versus [O\,{\sc iii}]$\lambda$5007/H$\beta$ 
diagram from \citet{kauffmannetal2003} outlines well the area occupied by staburst-like objects. 

We find also that two giant  H\,{\sc ii} regions located at different positions inside the disc (one 
 H\,{\sc ii} region may be associated with circumnuclear star formation) seem 
to be responsible for the double-peaked emission lines in the spectra of Sample A.  
However, two starbursts in two different galaxies which are projected on top of
each other, as alternative scenario for the origin of the double-peaked emission lines, 
cannot be excluded. Photometric and spectroscopic 
observations with higher resolution may help to resolve this issue.

\section*{Acknowledgments}


L.S.P. and I.A.Z acknowledge support from the Cosmomicrophysics project of
the National Academy of Sciences of Ukraine. 
Part of this work was supported by the Spanish Plan Nacional de Astronom\'{\i}a y 
Astrof\'{\i}sica under grants AYA2008-06311-C02-01,  AYA2007-67965-C03-02 and AYA2010-21887-C04-01.

Funding for the SDSS and SDSS-II has been provided by the Alfred P. Sloan Foundation, 
the Participating Institutions, the National Science Foundation, the U.S. Department of Energy,
the National Aeronautics and Space Administration, the Japanese Monbukagakusho, the Max Planck Society, 
and the Higher Education Funding Council for England. The SDSS Web Site is http://www.sdss.org/.

The SDSS is managed by the Astrophysical Research Consortium for the Participating Institutions. 
The Participating Institutions are the American Museum of Natural History, Astrophysical Institute Potsdam,
University of Basel, University of Cambridge, Case Western Reserve University, University of Chicago, 
Drexel University, Fermilab, the Institute for Advanced Study, the Japan Participation Group, 
Johns Hopkins University, the Joint Institute for Nuclear Astrophysics, the Kavli Institute for Particle 
Astrophysics and Cosmology, the Korean Scientist Group, the Chinese Academy of Sciences (LAMOST), 
Los Alamos National Laboratory, the Max-Planck-Institute for Astronomy (MPIA), the Max-Planck-Institute 
for Astrophysics (MPA), New Mexico State University, Ohio State University, University of Pittsburgh, 
University of Portsmouth, Princeton University, the United States Naval Observatory, and 
the University of Washington.

\end{document}